%% file: main.tex
\documentclass[]{spie}  %>>> use for US letter paper
\usepackage{subcaption}
\usepackage{amsmath,amsfonts,amssymb}
\usepackage{graphicx}
\usepackage[colorlinks=true, allcolors=blue]{hyperref}
\usepackage[leftcaption]{sidecap}

\title{Classification of Schizophrenia from Functional MRI Using Large-scale Extended Granger Causality}

\author[a,b,c,d]{Axel Wismüller,}
\author[a]{M. Ali Vosoughi}

\affil[a]{Department of Electrical and Computer Engineering, University of Rochester, NY, USA}
\affil[b]{Department of Imaging Sciences, University of Rochester, NY, USA}
\affil[c]{Department of Biomedical Engineering, University of Rochester, NY, USA}
\affil[d]{Faculty of Medicine and Institute of Clinical Radiology, Ludwig Maximilian University,
Munich, Germany}

\authorinfo{Further author information: (Send correspondence to Ali Vosoughi)\\Ali Vosoughi: E-mail: mvosough@ur.rochester.edu}

\begin{document} 
\maketitle

\begin{abstract}
The literature manifests that schizophrenia is associated with alterations in brain network connectivity. We investigate whether large-scale Extended Granger Causality (lsXGC) can capture such alterations using resting-state fMRI data. Our method utilizes dimension reduction combined with the augmentation of source time-series in a predictive time-series model for estimating directed causal relationships among fMRI time-series. The lsXGC is a multivariate approach since it identifies the relationship of the underlying dynamic system in the presence of all other time-series. Here lsXGC serves as a biomarker for classifying schizophrenia patients from typical controls using a subset of 62 subjects from the Centers of Biomedical Research Excellence (COBRE) data repository. We use brain connections estimated by lsXGC as features for classification. After feature extraction, we perform feature selection by Kendall's tau rank correlation coefficient followed by classification using a support vector machine. As a reference method, we compare our results with cross-correlation, typically used in the literature as a standard measure of functional connectivity. We cross-validate 100 different training/test (90\%/10\%) data split to obtain mean accuracy and a mean Area Under the receiver operating characteristic Curve (AUC) across all tested numbers of features for lsXGC. Our results demonstrate a mean accuracy range of [0.767, 0.940] and a mean AUC range of [0.861, 0.983]  for lsXGC. The result of lsXGC is significantly higher than the results obtained with the cross-correlation, namely mean accuracy of [0.721, 0.751] and mean AUC of [0.744, 0.860]. Our results suggest the applicability of lsXGC as a potential biomarker for schizophrenia.
\end{abstract}


\input{introduction}
\input{data}
\input{methods}

\input{results}
\input{conclusion}

\acknowledgments % equivalent to \section*{ACKNOWLEDGMENTS}       
 
This research was funded by Ernest J. Del Monte Institute for Neuroscience Award from the Harry T. Mangurian Jr. Foundation. This work was conducted as a Practice Quality Improvement (PQI) project related to American Board of Radiology (ABR) Maintenance of Certificate (MOC) for Prof. Dr. Axel Wismüller. This work is not being and has not been submitted for publication or presentation elsewhere.  

% References
\bibliography{report} % bibliography data in report.bib
\bibliographystyle{spiebib} % makes bibtex use spiebib.bst

\end{document}

%% file: introduction.tex
\section{INTRODUCTION} \label{sec:intro}  
Schizophrenia is a psychiatric disorder characterized by thoughts or experiences that are out of touch with reality,  decreased participation in daily activities, disorganized speech or behavior, and (probably) difficulty with concentration and memorization may also be present. The current diagnosis of schizophrenia is by using clinical evaluations of symptoms and behaviors; nevertheless, measurable biomarkers can be beneficial. Recent studies on brain imaging data have shown that information can be extracted non-invasively from brain activity.  Despite these studies' promising results, there is still scope for improvement, especially using more meaningful connectivity analysis approaches [\citeonline{li2020neuroimaging}].

Extensive evidence has demonstrated that schizophrenia affects the brain's connectivity [\citeonline{calhoun2012exploring}]. Biomarkers from resting-state functional MRI (rs-fMRI) for schizophrenia can be derived using Multi-Voxel Pattern Analysis (MVPA) techniques [\citeonline{4_norman2006beyond}]. MVPA is a framework based on pattern recognition that extracts differences in brain connectivity patterns among healthy individuals and individuals with neurological disease. Cross-correlation is commonly used in most MVPA studies to obtain a functional connectivity profile. For instance, one such study has obtained an accuracy of 0.79 on the slow frequency bands (0.01-0.1 Hz) [\citeonline{cheng2015nodal}]. As a result,  one can argue that connectivity analysis of fMRI data can be used to learn meaningful information. However, cross-correlation is not fit to obtain directed measures of connectivity. Therefore, there may be more relevant information in the fMRI data that is not being grasped by cross-correlation. Several methods have been proposed to capture directional relations in multivariate time-series data, e.g., transfer entropy [\citeonline{schreiber2000measuring}] and mutual information [\citeonline{kraskov2004estimating}]. However, as the multivariate problem's dimensions increase, the density function's computation becomes computationally expensive [\citeonline{mozaffari2019online,mozaffari2019online_ieee}]. Under the Gaussian assumption, transfer entropy is equivalent to Granger causality [\citeonline{barnett2009granger}]. However, the computation of multivariate Granger causality for short time series in large-scale problems is challenging [\citeonline{vosoughi2020large,dsouza2020large}]. 

Large-scale Extended Granger Causality (lsXGC) is a recently proposed method for estimating directed causal relationships among fMRI time-series that combines dimension reduction with source time-series augmentation and uses predictive time-series modeling   [\citeonline{vosoughi2021marijuana}]. In this work, we investigate if alterations in directed connectivity evident in individuals with schizophrenia and if such directed measures enhance our ability to discriminate between schizophrenia patients and healthy controls. To this end, we apply lsXGC in the MVPA framework for estimating a measure of directed causal interdependence between fMRI time-series. 
\newline
This work is embedded in our group’s endeavor to expedite artificial intelligence in biomedical imaging by means of advanced pattern recognition and machine learning methods for computational radiology and radiomics, e.g., [ \citeonline{nattkemper2005tumor,bunte2010adaptive,8_wismueller2000segmentation,9_leinsinger2006cluster,10_wismuller2004fully,11_hoole2000analysis,12_wismuller2006exploratory,13_wismuller1998neural,14_wismuller2002deformable,15_behrends2003segmentation,16_wismuller1997neural,17_bunte2010exploratory,18_wismuller1998deformable,19_wismuller2009exploration,20_wismuller2009method,22_huber2010classification,23_wismuller2009exploration,24_bunte2011neighbor,25_meyer2004model,26_wismuller2009computational,27_meyer2003topographic,28_meyer2009small,29_wismueller2010model,meyer2007unsupervised,30_huber2011performance,31_wismuller2010recent,meyer2007analysis,32_wismueller2008human,wismuller2015method,33_huber2012texture,34_wismuller2005cluster,35_twellmann2004detection,37_otto2003model,38_varini2004breast,39_huber2011prediction,40_meyer2004stability,41_meyer2008computer,42_wismuller2006segmentation,45_bhole20143d,46_nagarajan2013computer,47_wismuller2004model,48_meyer2004computer,49_nagarajan2014computer,50_nagarajan2014classification,yang2014improving,wismuller2014pair,51_wismuller2014framework,schmidt2014impact,wismuller2015nonlinear,wismuller2016mutual,52_schmidt2016multivariate,abidin2017using,61_dsouza2017exploring,53_chen2018mri,54_abidin2018alteration,55_abidin2018deep,dsouza2018mutual,chockanathan2019automated} ].

%% file: data.tex
\section{DATA} \label{sec:data}
\subsection{Participants}
The Centers of Biomedical Research Excellence (COBRE) data respiratory contains raw anatomical and functional MR data from 72 patients with schizophrenia and 74 healthy controls (ages ranging from 18 to 65 in each group). All subjects were screened and eliminated if they had; a history of mental retardation, a history of neurological disorder,  history of severe head trauma with more than 5 minutes of loss of consciousness, history of substance dependence or abuse within the last 12 months [\citeonline{calhoun2012exploring}]. Diagnostic information was collected using the Structured Clinical Interview used for DSM Disorders (SCID) [\citeonline{calhoun2012exploring}]. 

\subsection{Resting-state fMRI data}
A multi-echo MPRAGE (MEMPR) sequence was used with the following parameters: TR/TE/TI = 2530/[1.64, 3.5, 5.36, 7.22, 9.08]/900 ms, flip angle = $7^{\circ}$, FOV = 256 x 256 mm$^2$, slab thickness = 176 mm, matrix = 256 x 256 x 176, voxel size =1 x 1 x 1 mm$^3$, number of echos = 5, pixel bandwidth = 650 Hz, total scan time = 6 min. With 5 echoes, the TR, TI and time to encode partitions for the MEMPR are similar to that of a conventional MPRAGE, resulting in similar GM/WM/CSF contrast. Resting-state fMRI data was collected with single-shot full k-space echo-planar imaging (EPI) with ramp sampling correction using the intercomissural line (AC-PC) as a reference (TR = 2 s, TE = 29 ms, matrix size = 64 x 64, 32 slices, voxel size = 3 x 3 x 4 mm$^3$).

Functional connectivity measurements were generated from a subsample of the COBRE dataset [\citeonline{niak2016cobre}], a publicly available sample which we accessed through the Nilearn Python library [\citeonline{nilearn}]. All subjects of healthy controls and diseased patients under the age of 32 were selected, including 33 healthy and 29 diseased subjects, totaling 62 individuals. The images were already preprocessed using the NIAK resting-state pipeline [\citeonline{niak-preprocessed}], and additional details can be found in the reference [\citeonline{niak2016cobre}]. The number of regions of interests has been selected to be 122 with functional brain parcellations [\citeonline{bellec2013mining}]. 

% \donotshow{
% \subsection{Synthetic directional network}
% We examine the effectiveness of LsXGC to reliably estimate influence scores for the synthetic bidirectional networked time-series data with known ground truth \cite{baccala2001partial}. We analyzed the toy model number 5 

% \donotshow{ whose network structure is shown in Fig. \ref{fig:smith_sim4_PR}}, which has 50 nodes with 1000 time samples for each node. In this paper, we investigate LsaGC for time-series at two different signal-to-noise (SNR) levels of SNR=5, and 15 dB.
% }

%% file: methods.tex
\section{METHODS}\label{sec:methods}
\subsection{Large-scale Extended Granger Causality (lsXGC)}

The Large-scale Extended Granger Causality (lsXGC) method has been developed based on 1) the principle of original Granger causality that quantifies the causal influence of time-series $\mathbf{x_s}$ on time-series $\mathbf{x_t}$ by quantifying the measure of improvement in the forecast of $\mathbf{x_t}$ in the presence of $\mathbf{x_s}$. 2) the idea of dimensionality reduction, which solves the problem of tackling an ill-posed system, which is often challenged in fMRI analysis since the number of acquired temporal samples usually is not sufficient for estimating the model parameters [\citeonline{61_dsouza2017exploring, vosoughi2020large}].

Consider the ensemble of time-series $\mathcal{X}\in \mathbb{R}^{N\times T}$, where $N$ is the regions of interest (ROIs or number) of time-series and $T$ the number of temporal samples. Let $\mathcal{X} = (\mathbf{x_1}, \mathbf{x_2}, \dots, \mathbf{x_N})^{\mathsf{T}}$ be the whole multidimensional system and $x_i \in \mathbb{R}^{1\times T}$ a single time-series with $i = 1, 2, \dots,N$, where $\mathbf{x_i} = (x_i(1), x_i(2), \dots, x_i(T))$. To overcome the ill-posed problem, first $\mathcal{X}$ will be decomposed into its first $p$ high-variance principal components
$\mathcal{Z} \in \mathbb{R}^{p\times T}$ using Principal Component Analysis (PCA), i.e.,

\begin{equation}
\mathcal{Z}=W\mathcal{X},    
\end{equation}

where $W\in \mathbb{R}^{p\times N}$ represents the PCA coefficient matrix. Subsequently, the dimension-reduced time-series ensemble $\mathcal{Z}$ is augmented by one original time-series $\mathbf{x_s}$ yielding a dimension-reduced augmented time-series ensemble $\mathcal{Y}\in \mathbb{R}^{(p+1)\times T}$ for estimating the influence of $\mathbf{x_s}$ on all other time-series.

Following this, we locally predict $\mathcal{X}$ at each time sample $t$, i.e., $\mathcal{X}(t)\in \mathbb{R}^{N\times 1}$ by calculating an estimate $\hat{\mathcal{X}}_{\mathbf{x_s}}(t)$. To this end, we fit an affine model based on a vector of $m$ vector of m time samples of $\mathcal{Y}(\tau)\in \mathbb{R}^{(p+1)\times 1}$($\tau=t-1, t-2, \dots, t-m$), which is $\mathbf{y}(t)\in \mathbb{R}^{m.(p+1)\times 1}$, and a parameter matrix $\mathcal{A}\in \mathbb{R}^{N\times m.(p+1)}$ and a constant bias vector $\mathbf{b}\in \mathbb{R}^{N\times 1}$, 
\begin{equation}
    \hat{\mathcal{X}}_{\mathbf{x_s}}(t)=\mathcal{A}\mathbf{y}(t)+\mathbf{b},~~ t=m+1, m+2, \dots, T.
\end{equation}

Now $\hat{\mathcal{X}}_{\setminus {\mathbf{x_s}}}(t)$, which is the prediction of $\mathcal{X}(t)$ without the information of $\mathbf{x_s}$, will be estimated. The estimation processes is identical to the previous one, with the only difference being that we have to remove the augmented time-series $\mathbf{x_s}$ and its corresponding column in the PCA coefficient matrix $W$.

The computation of a lsXGC index is based on comparing the variance of the prediction errors obtained with
and without consideration of $\mathbf{x_s}$. The lsXGC index $f_{\mathbf{x_s}\xrightarrow{}\mathbf{x_t}}$ , which indicates the influence of $\mathbf{x_s}$ on $\mathbf{x_t}$, can be calculated by the following equation:
\begin{equation}
    f_{\mathbf{x_s}\xrightarrow{}\mathbf{x_t}}=\log {\frac{\mathrm{var}(e_s)}{\mathrm{var}(e_{\setminus s})}},
\end{equation}
where $e_{\setminus s}$ is the error in predicting $\mathbf{x_t}$ when $\mathbf{x_s}$ was not considered, and $e_s$ is the error, when $\mathbf{x_s}$ was used. In this study, we set $p = 8$ and $m = 1$.

\subsection{Multi-voxel pattern analysis}
Brain connections served as features for classification in this study and were estimated by two methods, namely lsXGC and cross-correlation. Before using high-dimensional connectivity feature vectors as input to a classifier, feature selection was carried out to reduce input features' dimension.

\subsubsection{Feature selection}
In order to lessen the number of features, feature selection was performed on each training data set with k-fold cross-validation using \textit{Kendall's Tau} rank correlation coefficient [\citeonline{62_kendall1945treatment}] and $10\%-90\%$ of test-to-train split ratio. This approach quantifies each feature's relevance to the task of classification and assigns ranks by testing for independence between different classes for each feature [\citeonline{62_kendall1945treatment}]. 

\subsubsection{Classification}
To cross-validate the classification performance in 100 iterations, the data set was divided into two groups: a training data set ($90\%$) and a test data set ($10\%$) that the percentage of samples for each class was preserved. Also, this was repeated with different numbers of features ranging from 5 to 175. A Support Vector Machine (SVM) [\citeonline{63_suykens1999least}] was used for classification between healthy subjects and schizophrenia patients. All procedures were performed using MATLAB 9.8 (MathWorks Inc., Natick, MA, 2020a), and Python 3.8.

%% file: results.tex
\section{RESULTS}\label{sec:results}
\begin{figure}
    \centering
    \includegraphics[width=\textwidth]{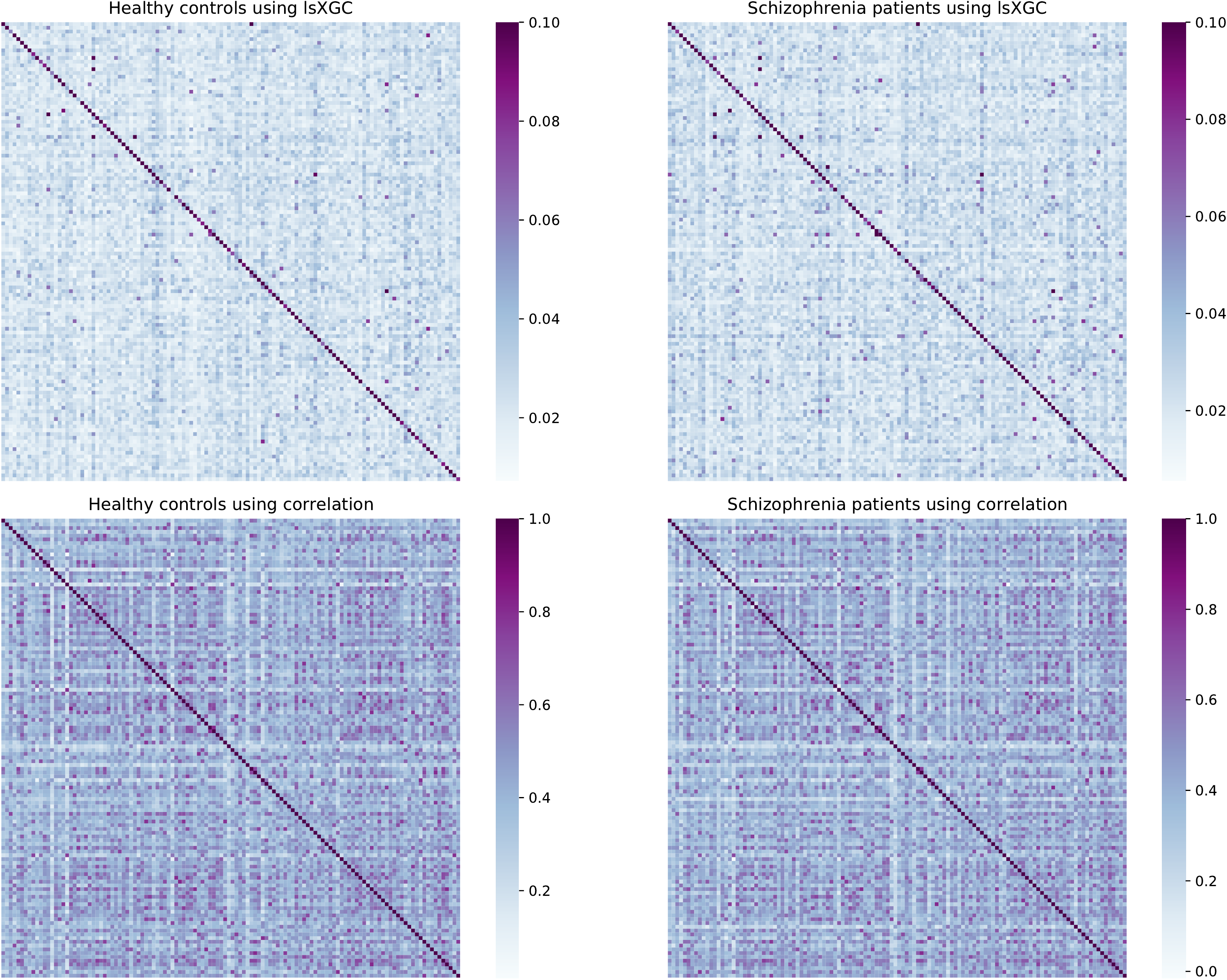}
    \caption{Mean connectivity matrices: top left: mean connectivity of healthy control subjects using lsXGC, top right: mean connectivity matrix of schizophrenia patients using lsXGC, bottom left: mean connectivity matrix of healthy control subjects using cross-correlation, bottom right: employing cross-correlation to obtain mean connectivity matrix of schizophrenia patients. Remarkably different methods appear to extract different connectivity features, and that they appear to be slight differences in connectivity patterns between the healthy subject and the schizophrenia patients. }
    \label{fig:matrix_plot}
\end{figure}

Mean connectivity matrices, which were extracted using lsXGC and cross-correlation, are shown in Fig. \ref{fig:matrix_plot} for schizophrenia patient and healthy control cohorts. Distinct patterns are visible to the naked eye for both methods. In the following, we quantitatively investigate the difference between the two patient cohorts' connectivity patterns using an MVPA approach. 

% \vspace{-.1in}

Classification results were evaluated using the Area Under the Receiver Operator Characteristic Curve (AUC) and accuracy. An AUC = 1 indicates a perfect classification, AUC = 0.5 indicates random class assignment.
\begin{figure}
     
     \centering
     \begin{subfigure}[b]{0.4\textwidth}
         \centering
         \includegraphics[width=\textwidth]{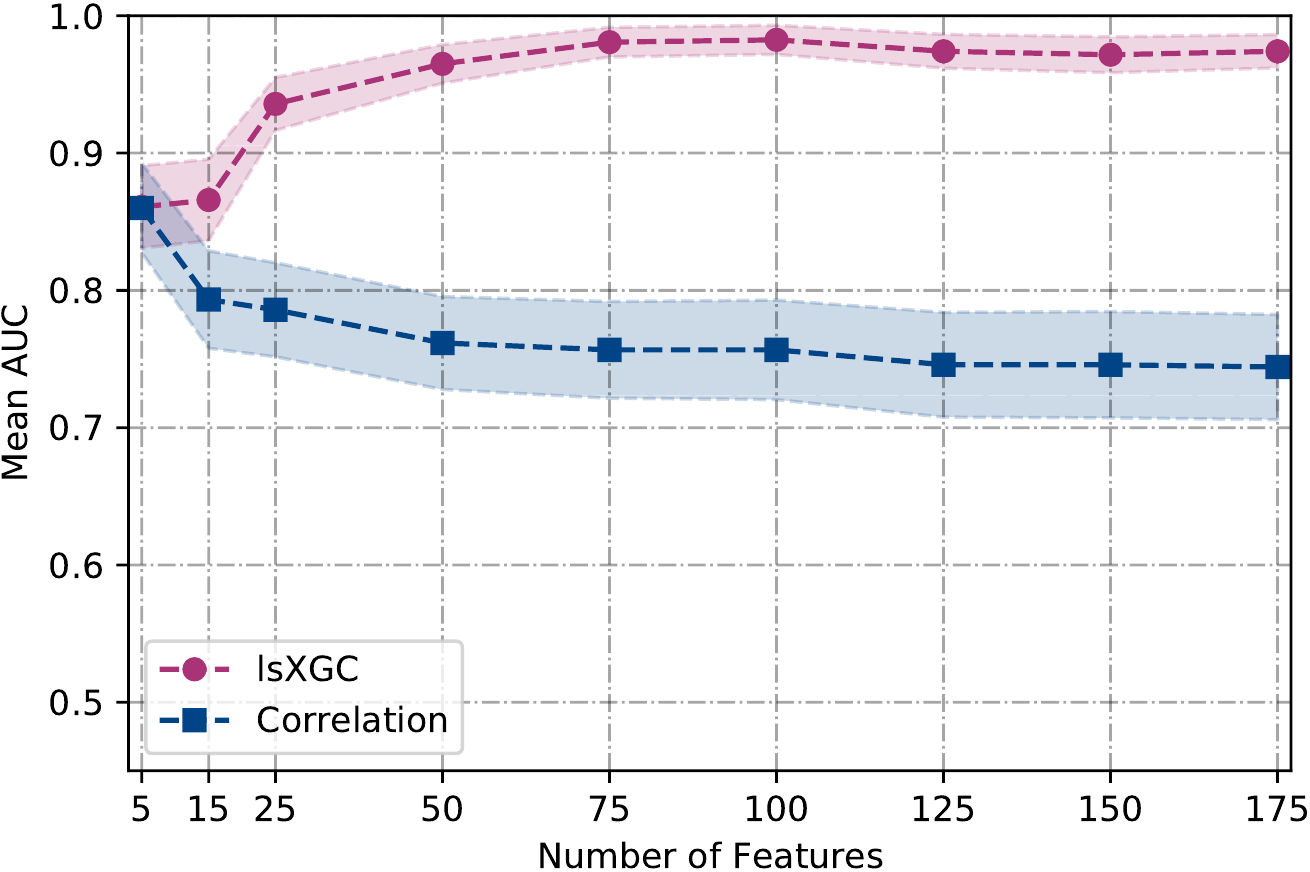}
         \caption{Mean AUC}
         %\label{fig:three sin x}
     \end{subfigure}
     \begin{subfigure}[b]{0.4\textwidth}
         \centering
         \includegraphics[width=\textwidth]{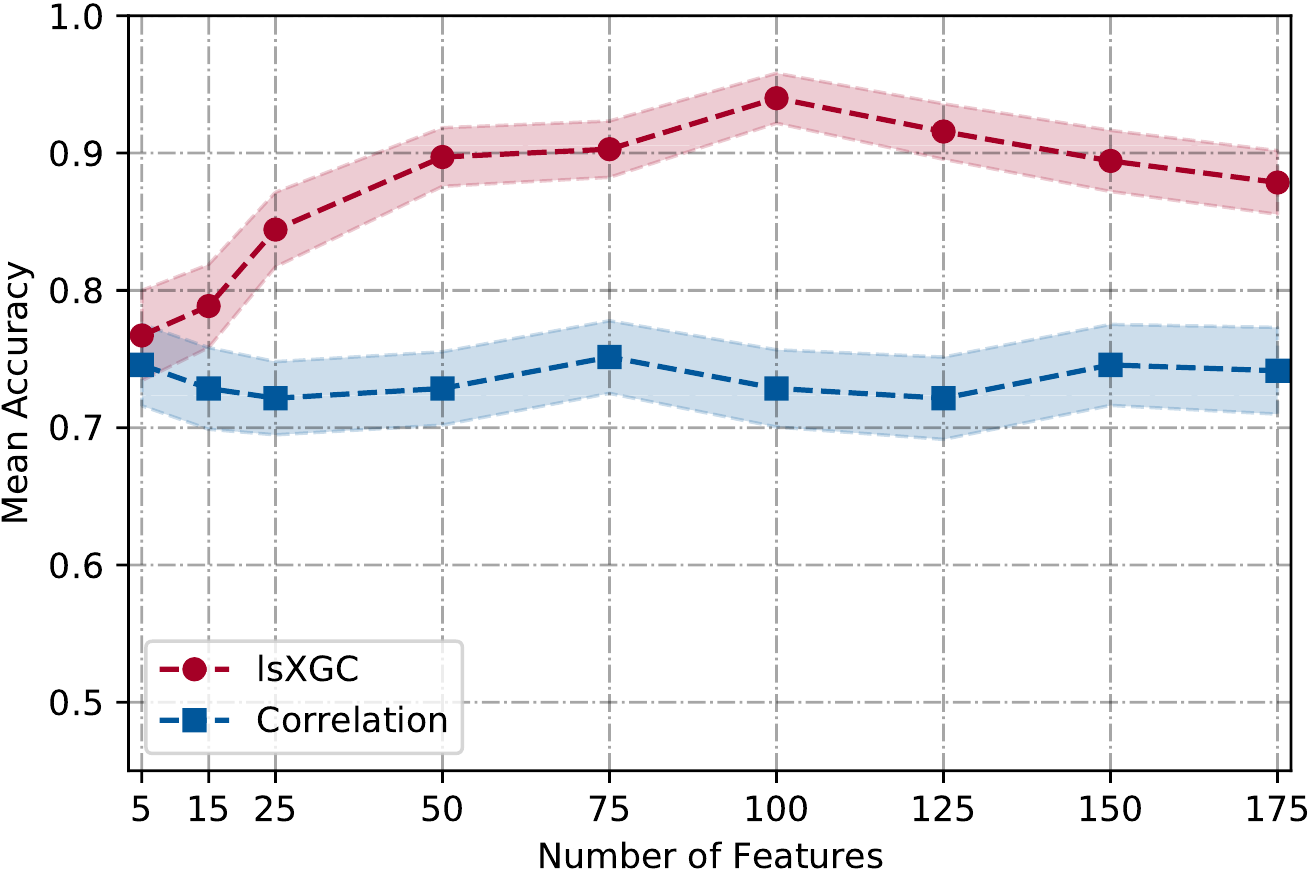}
         \caption{Mean accuracy}
         %\label{fig:five over x}
     \end{subfigure}
     \vspace{.1in}
        \caption{Plots are comparing the performance of cross-correlation and the proposed large-scale extended Granger causality (lsXGC). The shaded areas represent the $95\%$ confidence interval. It demonstrates that lsXGC outperforms cross-correlation for most numbers of selected features.}
        \label{fig:auc_plots}
\end{figure}
In this study, we chose eight as the number of the retained components of PCA in the lsXGC algorithm and model order of 1 for the multivariate vector autoregression function based on preliminary analyses. The plots of accuracy and AUC results in Fig. \ref{fig:auc_plots}, clearly demonstrate that lsXGC outperforms cross-correlation for diversified numbers of features. Across the wide range of examined numbers of features, the performance of lsXGC is consistently higher with its mean AUC within [0.861, 0.983] and its mean accuracy within [0.767, 0.940]. On the other hand, cross-correlation performs quite poorly compared to lsXGC with its mean AUC within [0.744, 0.860] and its mean accuracy within [0.721, 0.751].

%% file: conclusion.tex
\section{CONCLUSIONS}\label{sec:conclusions}
In this research, we use a recently developed method for brain connectivity analysis, large-scale Extended Granger Causality (lsXGC), and apply it to a subset of the COBRE data repository to classify individuals with schizophrenia from typical controls by estimating a measure of directed causal relations among regional brain activities recorded in resting-state fMRI. Following the construction of connectivity matrices as characterizing features for brain network analysis, we use Kendall's tau rank correlation coefficient to select a significant feature and a support vector machine to classify. We demonstrate that our method (lsXGC) favorably compares to standard analysis using cross-correlation, as shown by the significantly enhancing accuracy and AUC values. The effectiveness of lsXGC as a potent biomarker for identifying schizophrenia in prospective clinical trials is yet to be validated. Nevertheless, our results suggest that our approach outperforms the current clinical standard, namely cross-correlation, at revealing meaningful information from functional MRI data. 

% \donotshow{We further quantified the performance of the lsXGC method as compared to cross-correlation using the synthetic network with known ground truth, and we demonstrate that lsXGC captures the directional patterns of interactions compared to cross-correlation, as indicated by the significantly higher AUC under various noise values.}